# Deep-learning-based optical image hiding

**Jiaosheng Li,[2,§] Yuhui Li,[1,§] Ju Li,[1,§] Qinnan Zhang,[3] Guo Yang,[1] Shimei Chen,[1] Chen Wang[4] and Jun Li [1,*]**

[1] *Guangdong Provincial Key Laboratory of Quantum Engineering and Quantum Materials, School of Physics and Telecommunication Engineering, South China Normal University, Guangzhou 510006, China*
[2] *Guangdong Provincial Key Laboratory of Nanophotonic Functional Materials and Devices, South China Normal University, Guangzhou 510006, China*
[3] *Shenzhen Key Laboratory of Micro-Nano Measuring and Imaging in Biomedical Optics, College of Optoelectronic Engineering, Shenzhen University, Shenzhen 518060, China*
[4] *School of Electronic Science and Engineering, Nanjing University*

*§These authors contributed equally to this work*
*\*Corresponding author:lijunc@126.com*

**Abstract:** A novel framework of optical image hiding based on deep learning (DL) is proposed in this paper, and hidden information can be reconstructed from an interferogram by using an end to end network with high-quality. By using the prior data between the hidden image and the object image, a generative adversarial network was trained so that it can learn the hiding model, which resulting in only an interferogram needs to be transmitted and recorded to reconstruct image. Moreover, reconstruction process can be obtained without the parameters in optical inverse diffraction and the reconstruction result will not be affected by the phase shifts deviation and noise, which is convenient for practical application. The feasibility and security of the proposed method are demonstrated by the optical experiment results.

## 1. Introduction

Deep learning (DL) obtains the approximation of the optimal model of the system through large amount of prior information. In the real imaging or security system, we can use the prior information to obtain more accurate parameters of the model through the computer. Based on this, DL has shown its advantages in many applications, such as optical imaging [1-3], phase recovery and holographic image reconstruction [4, 5], computational ghost imaging [6], Cryptanalysis [7,8], etc. [9]. Convolutional networks [10] shows its powerful function in optical image reconstruction [1, 3, 11], and can be used to reconstruct the object wavefront directly from a single-shot in-line digital hologram [5].

Optical image security technology has become an important research direction in the field of information security because of its advantages of high efficiency, high security, difficult to crack and easy to store [12-14]. Since the double random phase encoding (DRPE) technology was put forward [15], optical image security technology has developed rapidly by introducing additional degrees of freedom, such as wavelength, propagation distance [15], degree of polarization [16]. By embedding the original information into the host information without destroying the form of the original host information, optical image hiding can achieve the purpose of hiding information and not attracting the attention of the attacker [17-20], thus reducing the possibility of being violated. On this basis, the classical methods in cryptography are used to enhance the security of the hidden information. Optical image hiding technology based on phase-shifting interferometry (PSI) [21-24] can not only make use of the spatial bandwidth product of CCD to record the interferogram information, but also make up the shortage of in-line and off-axis digital holography. However, it needs to get multiple phase-shifting interferograms with specific phase shifts and the usage of various optical encryption keys to realize image reconstruction. The amount of data transmitted is large, and the quality of decrypted image is easily affected by system error and phase shifts deviation, which is difficult to be applied in practice.

Based on this, an optical image hiding method based on deep learning (DL) is proposed in this paper. By using the prior information between the hidden image and the object image, the proposed method can obtain the approximation of hiding model based on the optical inverse diffraction through the DL method. Based on the generative adversarial network (GAN) [25], by setting the corresponding constraints, that is, the hidden image as the constraint of DL, the conditional generative adversarial network (CGAN) [26] can continuously learn the corresponding relationship between the hidden image and the reconstructed image, so that only one hidden interferogram is needed to realize the reconstruction of optical image hiding. In the proposed method, the optical system no longer needs phase-shifting device, and the reconstruction process can be obtained without the parameters in optical inverse diffraction and the reconstruction result will not be affected by the phase shifts deviation and noise. Besides, only an interferogram needs to be transmitted, so the transmission efficiency is greatly improved. In this paper, the feasibility and superiority of the method are verified by the experiment, and the security, generalization, image reconstruction results under different data volumes are analyzed in detail.

## 2. Method analysis

Here, a DL-based image hiding method is introduced. By using the prior data between the hidden image and the object image, the hiding model based on optical diffraction is obtained by DL, and the decryption process can be obtained without the parameters in optical inverse diffraction. The schematic of the DL-based image hiding method is shown in Fig. 1. A series of hidden images are collected by using the optical hiding experiment system, and then the original image and the corresponding hidden image are input to the PC end to train by the CGAN. After the training, the optimized generation network can be obtained to reconstruct the object image from hidden interferogram directly.

### 2.1 Review of optical image hiding based on modified Mach-Zehnder interferometer

The basic principle of optical image hiding will briefly be introduced in this section. The schematic of optical image hiding is showed in Fig.1. The laser beam is divided into an object beam and a reference beam. The object beam first irradiates the target image on the spatial light modulator (SLM); the host image is placed in the reference beam to achieve the modulation of the reference beam.

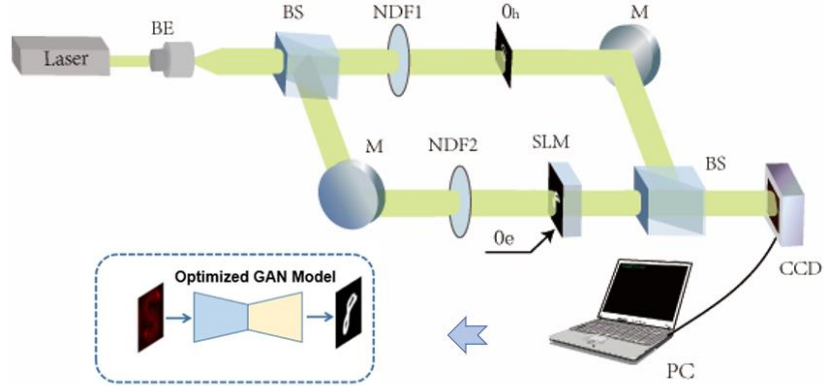

Fig.1. The Schematic of the DL-based image hiding method. BE, beam expander; BS, beam splitter; M, mirror; NDF, neutral density filters; Oe, object image; Oh, host image.

Assuming that the transmissivity of the original image can be expressed as $O(x_0, y_0)$, then the complex amplitude of the original image on the CCD plane can be expressed as:

$$\psi_0(x, y) = Frt[O(x_0, y_0)] = A(x, y)\exp[i\phi(x, y)] \quad (1)$$

Where $Frt$ represents the Fresnel transform of $O(x_0, y_0)$, then the complex amplitude distribution of the host image can be expressed as:

$$\psi_h(x, y) = A_h(x, y)\exp[i\phi_h(x, y)] \quad (2)$$

Therefore, the interference intensity distribution obtained on the CCD can be written as:

$$\begin{aligned} I(x, y) &= |\psi_0(x, y) + \psi_h(x, y)|^2 \\ &= A(x, y)^2 + A_h(x, y)^2 \\ &\quad + 2A(x, y)A_h(x, y)\cos[\phi_h(x, y) - \phi(x, y)] \end{aligned} \quad (3)$$

In the process of image hiding, the original image is embedded in the diffraction field of the host image. Different hiding effects can be achieved through adjusting the light intensity ratio of the object beam and the reference beam by adjusting two neutral density filters in the experimental system. Therefore, the interferogram can look like the diffraction field of the host image when original image is hidden. The traditional reconstruction method needs to add a phase-shifting device in the experimental system to collect at least two-frame phase-shifting interferograms of hidden image and host image, and then use the multi-step phase-shifting method to reconstruct them. In the proposed method, the experimental data set can be built by continuously loading different original images on SLM, and a CCD is used to record the resulting interference fringes.

### 2.2 The designed CGAN architecture

In our scheme, a CGAN is developed to train the object images (training label) and the corresponding hidden images (training data). The network structure diagram of CGAN is shown in Fig.2, which consists of two parts, i.e. generator network and discriminating network. The hidden image is firstly entered the generator network to get the estimated image of the corresponding object image, then the estimated image and the hidden image will form a "fake image pair" and the hidden image and the real object image form a "real image pair". These two pairs of images are input into the discrimination network and the corresponding discrimination matrix are generated, respectively. Then the loss of the discriminators is calculated to optimize the parameters of the discrimination network. At the same time, the discriminator loss is also involved in the optimization of the parameters of the generation network, so as to continuously optimize the network model. Finally, when the discrimination network cannot distinguish the generated image (that is, the generated image is very close to the label image), the DL model used for retrieving the object image from the hiding interferogram is learned by the generated network.

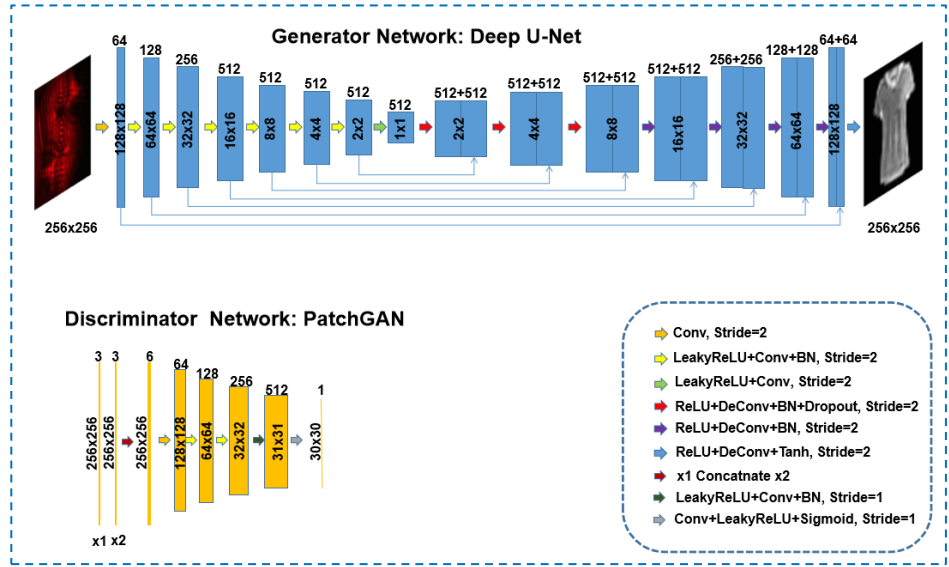

Fig.2. The designed CGAN architecture, including generator network and discriminator network.

In the diagram of generator network, the input and output of generator network are set to be hidden image and the object image. The hidden image firstly passes through 8 downsampling convolution layers to get the feature maps of hidden image, and then the estimate of the object image is produced by passing through 8 upsampling convolution layers. In addition, the downsampling layer and the upsampling layer of the generator can supplement the high-frequency information of the reconstructed object image through the skip connections. The discrimination network adopts PatchGAN structure [26], as shown in Fig. 2, and the input is "real image pair" or "fake image pair". The "image pair" is first concatenated by channels, and then a distinguish matrix with the size of $30 \times 30$ is obtained by five downsampling layers. In addition, the Dropout (p= 0.5) is introduced to the first three upsampling layers of the generator network to prevent over fitting. After training, the trained DL model can be used to retrieve the object image from the hiding interferogram. Here, the designed CGAN architecture is implemented by Python version 3.7 on a PC with Nvidia Geforce GTX1080Ti GPU.

### 3. Experimental results and discussions

The experimental setup is schematically showed in Fig. 1. A He-Ne laser (REO/30989) with the wavelength of 633 nm is used as illumination source. After the laser beam is expanded, the object beam and the reference beam modulated by the host image are respectively obtained through beam splitter (BS) to produce the hidden interferogram on CCD plane. Where SLM (FSLM07U-A) is used to load the object image, and a letter of 'S' is used as the host image. The images used as object image are handwritten-digit patterns from the MNIST database [27] and fashion-MNIST database (8-bit grayscale images with 28×28 pixels) [28], which are widely used for DL. The resulting interference fringes are recorded by a CCD (AVT Pike F-421B/C) with the size of 2048×2048 pixels, and the pixel pitch is 7.4 um. By loading a series of object images on SLM, the corresponding hidden image can be collected by CCD. Instead of making great effort to retrieve the object light wave by using multi-step phase-shifting method with various security keys, the object image can be retrieved directly by using a trained machine-learning model. 3000 images are loaded on the SLM to perform hiding and 2700 of the corresponding hidden images are used as the training data and the rest 300 as the testing data. The central 256×256 pixels of them are chosen to perform the training.

First, a set of hidden images embedded by handwritten-digit patterns from the MNIST database are used for training, the retrieved results are showed in Fig.3. Where the first row is the hidden interferograms, the corresponding ground truth and the reconstructed images are showed in the second and third row, respectively. Evidently, only one interferogram is needed without the usage of various optical encryption keys to retrieve the object images, and the retrieved images are of high quality compared with the ground truth. Moreover, the noise in the experimental system has almost no effect on the quality of the reconstructed image.

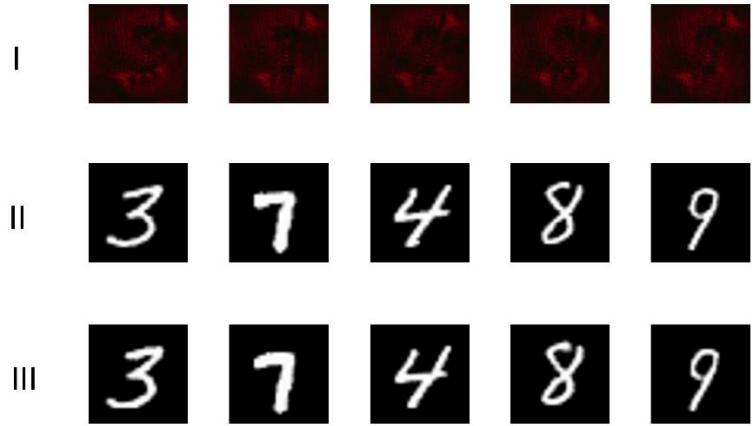

Fig3. Retrieved images of MNIST test data: (i) 5 testing hidden images from 300 testing data, (ii) ground truth (original object images), (iii) the retrieved object images by trained CGAN based-hiding method.

In order to further verify the feasibility of the method, 2700 fashion-MNIST database are used for training, in which the image content is more complex and the details are more abundant, and the retrieved results are distributed as shown in Fig. 4. The hidden interferograms, corresponding ground truth and the reconstructed images are showed in the first, second and third row of Fig.4. It can be seen from the collar and the pattern of the clothes in the retrieved images, the retrieved images are of high quality, in which the retrieved information can be visualized and the high-frequency features are retained.

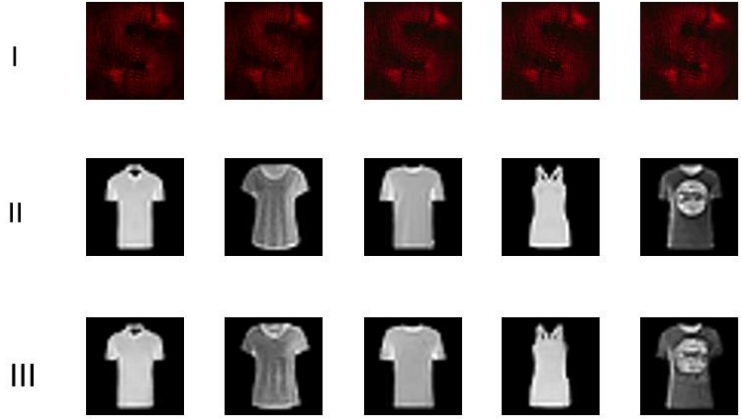

Fig.4. Retrieved images of fashion-MNIST test data: (i) 5 testing hidden images from 300 testing data, (ii) ground truth (original object images), (iii) the retrieved object images by trained CGAN based-hiding method.

We also analyze the reconstructed results of fashion-MNIST test data with different data volumes. Under the same parameter setting, 1000, 2000 and 3000 experimental pictures are applied respectively, and then predicted by using the same test images. The reconstructed results are shown in Fig. 5. Where the hidden images and the corresponding ground truth are showed in the first and second row, and the reconstructed images with different data volumes are showed in the last three rows, respectively. To quantify the performance of the reconstructed image under different data volumes during the training process, the correlation coefficient (CC) is introduced and can be defined as:

$$C = \mathrm{cov}(f, f_0)(\sigma_f, \sigma_{f_0})^{-1} \qquad (4)$$

where $f$ and $f_0$ denote the intensity distribution of reconstructed image and ground truth, respectively; and $\mathrm{cov}(f, f_0)$ is the covariance between $f$ and $f_0$, $\sigma_f$ and $\sigma_{f_0}$ are the standard deviations of $f$ and $f_0$, respectively. The CC values between the reconstructed images and the ground truth under different data volumes are plotted in Fig.6, in which the data volume changes from 1000 to 3000 with an interval of 1000. It can be seen from the results that with the increase of the number of data sets, the CC values and the quality of reconstructed image are improving accordingly. This shows that the quality of image restoration is directly related to the number of training data sets, but due to the limitations of experimental conditions and the time-consuming process of data set acquisition, we only collected 3000 experimental images for training, and the improvement of image quality in the later stage can be obtained through the increase of data volumes.

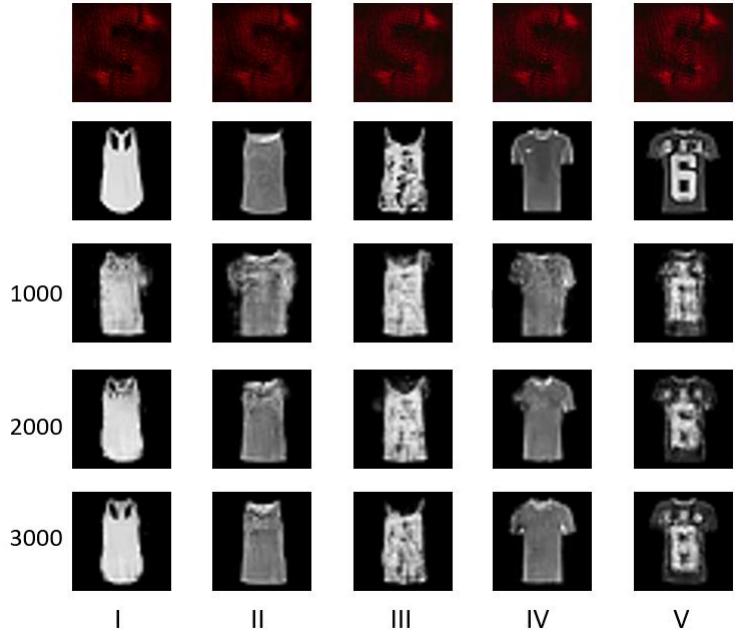

Fig.5. Retrieved images under different data volumes.

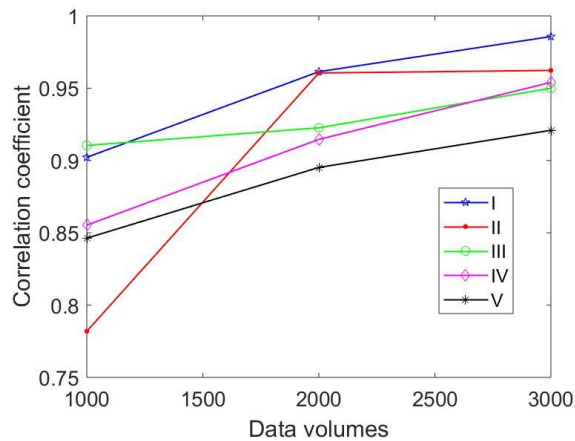

Fig.6. Correlation coefficient values under different data volumes.

In addition, to examine the generalization of the trained neural network model, here, we use it to recover the object images that are different from those in the training set. For convenience, we test it with the MNIST handwritten data set and the reconstructed images are shown in Fig. 7. The hidden images embedded by the MNIST database is showed in the first row, and the second and third row are the corresponding ground truth and the reconstructed images, respectively. Although the network was trained by using the fashion-MNIST test data, it is clearly seen that it can be used to reconstruct the different object images from MNIST handwritten data set, which shows that CGAN network can obtain the mathematical model of image reconstruction by DL well, and different object images hidden in the optical image hiding system can be reconstructed according to the optimized reconstruction network. However, due to the limited experimental data, the reconstructed image is a little fuzzy and the high-frequency details are lost, which can improve the performance by increasing the amount of data in the training set.

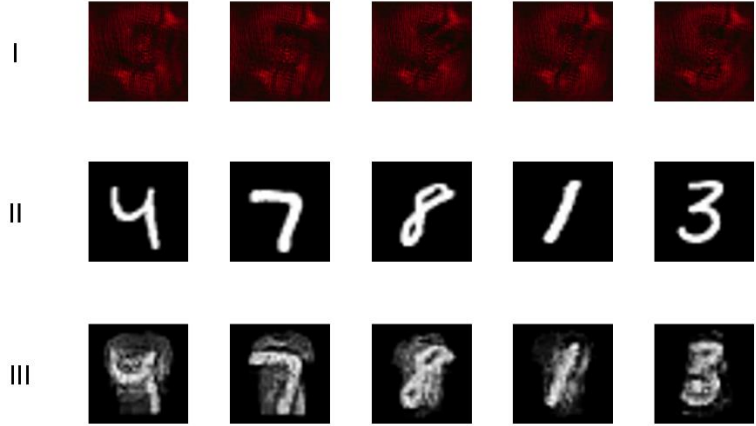

Fig.7. Network generalization test results. (i) hidden images embedded by MNIST handwritten data, (ii) ground truth (original object images), (iii) the object images retrieved form the network trained by fashion-MNIST test data.

At last, the security of the proposed method is discussed in this section. Keeping the original image reconstruction network trained by CGAN with 2700 training data unchanged (the letter 'S' is used as the host image), the host image of the optical image hiding system is changed to 'C', and the recorded hidden image from MNIST test data and fashion-MNIST test data are input into the CGAN-based reconstruction network. The reconstruction results are shown in Fig. 8, in which the first row is the hidden interferograms, the corresponding ground truth and the reconstructed images with the wrong key are showed in the second and third row, respectively. The results indicate that when the host image information is missing, the correct object information cannot be obtained completely, which further proves that proposed method has high security, and only the host image information in the optical image hiding system needs to be changed to resist the attack of DL-based system.

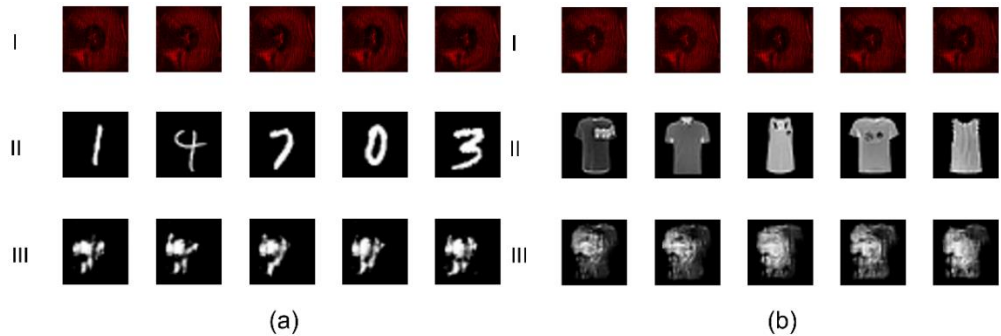

Fig.8. Retrieved images with incorrect host information. (a) MNIST test data, (b) fashion-MNIST test data.

## 4. Conclusion

In this paper, an optical image hiding method based on deep learning (DL) is proposed, in which the CGAN is used to train the physical model of the hidden system, and finally the physical model of the system is approached. In the proposed method, only a single interferogram is needed to achieve high-quality decryption, the transmission efficiency is greatly improved. And the optical system does not need the phase-shifting device, the decryption result will not be affected by the phase shifts deviation and noise, which is convenient for practical application. The feasibility and superiority of this method are verified by the experimental results, which reveals that the proposed method can provide an effective means to realize real-time and variable optical hiding embedding and extraction in the all-optical network and resist DL attacks.

**Funding**